\documentclass[a4paper,11pt]{article}
\usepackage{jheppub} 
\usepackage{lineno}
\usepackage{graphicx,amsfonts,amsmath,amssymb,epsfig,color, mathrsfs,latexsym,url,wasysym,float,xfrac, hyperref, appendix}
\usepackage[compatibility=false]{caption}
\usepackage{subcaption}

\newcommand{\be}{\begin{equation}}
\newcommand{\ee}{\end{equation}}
\newcommand{\nn}{\nonumber}
\newcommand{\ba}{\begin{eqnarray}}
\newcommand{\ea}{\end{eqnarray}}

\newcommand{\beq}{\begin{equation}}  \newcommand{\eeq}{\end{equation}}

\title{\boldmath Symmetry Points of ${\cal N}=1$ Modular Geometry}

\author{Amineh Mohseni,}
\author{Cumrun Vafa}

\affiliation{Jefferson Physical Laboratory, Harvard University, Cambridge, MA 02138, USA}

\emailAdd{amohseni@g.harvard.edu, vafa@g.harvard.edu}

\abstract{We consider 4d ${\cal N}=1$ supergravity theories with modular symmetry, where the modulus $\tau$ is the upper half-plane modulo $SL(2,\mathbf{Z})$ action. We focus on enhanced discrete gauge symmetry points $\tau=i, \exp(2\pi i/3)$, and argue that, if there are no new additional massless fields at these points, they will always be critical points of the scalar potential. Moreover, we show that whether these correspond to dS, AdS, or Minkowski vacua can be generically determined simply by the weight of the superpotential under modular transformations. We also analyze the asymptotics of the scalar potential and find that compatibility with the Swampland principles implies that, if nonvanishing, the scalar potential decays either exponentially or double-exponentially, and that the asymptotic slope is bounded. The slope is governed by the superpotential weight as well as by real-analytic modular contributions to the Kähler potential.
}

\begin{document}
\maketitle
\flushbottom
\section{Introduction}
String theory has led to a vast landscape of consistent supersymmetric models in four dimensions. However, despite great efforts over decades, reliable solutions to string theory without supersymmetry remain among the most challenging problems in constructing a complete model of our universe. Moreover, these efforts have motivated the belief that there are no \emph{stable} non-supersymmetric vacua in string theory. This, in turn, has led to attempts to construct non-supersymmetric metastable vacua (see \cite{McAllister:2023vgy} for a review). However, the reliability of these constructions has also been questioned \cite{Bena:2012tx,Bena:2012ek,Moritz:2017xto,Danielsson:2018ztv,Gautason:2018gln,Gao:2020xqh,Junghans:2022exo,Gao:2022fdi,Lust:2022lfc}. In addition, a general criterion, the Trans-Planckian Censorship Conjecture (TCC), has been proposed that provides obstructions to long-lived metastable non-supersymmetric dS vacua \cite{Bedroya:2019snp,Bedroya:2020rac,Bedroya:2020rmd,Bedroya:2022tbh}. Thankfully, these ideas are not in tension with the cosmology of our universe and have led to predictions of varying dark energy and a dark-matter sector \cite{Agrawal:2019dlm}; this model \cite{Bedroya:2025fwh} provides the best fit to the latest data from DES \cite{DES_SN5YR_2024,DES_Y3_Clusters_3x2pt_2025} and DESI \cite{DESI_DR2_BAO_2025,DESI_DR2_LyA_2025,DESI_DR2_ExtendedDE_2025}.
In particular, as the results of \cite{Bedroya:2025fwh} show, to match observations it is sufficient to be in a region of a non-supersymmetric landscape where $|\nabla V|\sim V$. This is relatively easy to satisfy, for example, near unstable critical points with $\nabla V=0$, by moving slightly away from those points. To complement these ideas, concrete constructions of unstable dS vacua from string theory have recently been achieved \cite{Chen:2025rkb,ValeixoBento:2025yhz}, showing that there is no obstruction to finding such vacua. Specifically, \cite{Chen:2025rkb} uses duality symmetries to argue for the existence of isolated (unstable) dS and AdS vacua. The method followed there was to use non-geometric flux vacua with 4d ${\cal N}=1$ supersymmetric EFTs and to search for unstable dS points at duality symmetric points in moduli.

One of our objectives in this paper is to study the extent to which duality symmetries can help identify AdS, dS, and Minkowski vacua in four-dimensional $\mathcal N=1$ supergravity within the string landscape. The advantage of looking for a non-supersymmetric vacuum starting from a supersymmetric EFT is that, although we lack the tools to compute the exact K\"ahler potential $K$ for these theories, aspects of the allowed potential are dictated by the holomorphic structure of the superpotential $W$. This places constraints on what is allowed and is sometimes under better analytic control.  Unfortunately, even that is not easy: despite progress in the computation of parts of the superpotential in various string constructions \cite{Witten:1996bn,Donagi:1996yf,Curio:1997rn,Gendler:2022qof}, we still do not have a single example for which we can compute the full non-perturbative superpotential.\footnote{For example for F-theory on CY elliptic 4-folds the effects of spacetime filling D3 branes or fluxes have not been fully incorporated into the superpotential computation.} Thus we have very little to start from in terms of concrete stringy examples!

Motivated by these partial stringy computations, we consider an ${\cal N}=1$ supersymmetric toy model with a single modulus $\tau$ parameterized by the upper half-plane, modulo $SL(2,{\bf Z})$ duality symmetries. Modular invariance requires the superpotential and Kähler potential to transform so that the generating functional \(G = K + \log\!\big(W\bar W\big)\) is invariant. Consequently, the superpotential is built from modular forms. That potentials will automatically have critical points at enhanced gauge symmetries is an old idea (see, e.g., \cite{Ginsparg:1986wr}). This idea has also been noted in the context of modular geometry at the symmetry points\footnote{See the caveat discussed later in this paper.} \cite{Font:1990gx, Cvetic:1991qm, Kokorelis:2000yt,Gonzalo:2018guu}. Modular properties were employed in \cite{Leedom:2022zdm} to study the scalar potential of four-dimensional heterotic toroidal orbifolds, and in \cite{Higaki:2024pql, Funakoshi:2024yxg} to investigate CP violation via moduli stabilization. Discrete symmetries were also used in \cite{Lust:2022mhk,Grimm:2024fip} to argue for the criticality and stabilization of the scalar potential in flux compactifications. With a rather different goal, this setup has also been recently studied in \cite{Casas:2024jbw,Kallosh:2024ymt,Kallosh:2024kgt,Carrasco:2025rud}.
 
 Specifically, we apply Swampland ideas as well as positivity properties of the kinetic term to place constraints on what is allowed. Assuming no extra massless modes at special points, discrete gauge symmetries enforced by duality symmetries automatically lead to critical points of the potential $V$ at the loci of enhanced discrete gauge symmetry, namely $\tau=i$ and $\tau=\exp(2\pi i/3)$. We find that, this yields (unstable) dS, supersymmetric AdS, and supersymmetric Minkowski solutions that can be predicted solely from the modular weight of the superpotential. Moreover, we analyze the asymptotic behavior of the scalar potential. We constrain the superpotential by imposing, for consistency with quantum gravity, that asymptotic blow-ups are forbidden. We also invoke swampland constraints—namely, the asymptotic gradient bound \cite{Bedroya:2019snp, Rudelius:2021oaz} \(
\left|\tfrac{\nabla V}{V}\right| \ge \sqrt{2},
\) to constrain the slope of the scalar potential. This slope is determined by the asymptotic Kähler potential, which can deviate from its tree-level value due to real-analytic corrections. Asymptotic double–exponential decay cannot be ruled out and, in fact, can potentially arise from gauge-instanton effects.

The organization of this paper is as follows: In Section~2 we introduce the basic setup and, in particular, the properties that modular invariance imposes on $W$ and $K$. In Section~3, we discuss the superpotential in greater detail and then investigate the resulting asymptotic structure of the theory in the weak-coupling regime, where $\tau \to \infty$. In Section~4 we explain how this framework leads to predictions of dS, AdS, and Minkowski vacua based on the modular weight at enhanced gauge-symmetry points. In Section~5 we end with some concluding thoughts.

\section{Modular Invariant $\mathcal{N}=1$ Theory}
\subsection{$\mathcal{N}=1$ SUGRA with Modular Symmetry}
Consider an $\mathcal{N}=1$, four-dimensional supergravity theory with $SL(2,\mathbb{Z})$ duality symmetry acting on a single modulus $\tau=\tau_1+i\tau_2$.
The modular group acts on the upper half-plane $\mathfrak{H}=\{\tau\in\mathbb{C}:\Im \tau>0\}$ by
\be
\gamma\cdot \tau=\frac{a \tau+b}{c \tau+d}
\qquad\text{for }\gamma=\begin{pmatrix}a&b\\ c&d\end{pmatrix}\in SL(2,\mathbb{Z}).
\ee
A fundamental domain for \(\Gamma\) is
\be
\mathcal{F}=\Bigl\{\tau\in\mathfrak{H}:\ |\tau|\ge 1,\ -\tfrac12\le \Re \tau\le \tfrac12\Bigr\}.
\ee
Two standard generators are
\be
S=\begin{pmatrix}0&-1\\ 1&0\end{pmatrix}\!: \tau\mapsto -\frac{1}{\tau},
\qquad
T=\begin{pmatrix}1&1\\ 0&1\end{pmatrix}\!: \tau\mapsto \tau+1.
\ee
The $SL(2,\mathbb{Z})$ symmetry, which can be viewed as a gauge symmetry, is broken at all points of moduli, except a discrete subgroup survives at two points. These two points with enhanced discrete gauge symmetry which we shall refer to as \textit{symmetry points} in the fundamental domain are: \(\tau=i\), fixed by \(S\) with $\mathbb{Z}_4$ gauge symmetry, and \(\tau=\omega=e^{2\pi i/3}\) with $\mathbb{Z}_6$ gauge symmetry, fixed by \(ST\). These have orders \(2\) and \(3\) left-over action on $\tau$-plane, respectively. $\tau \to \infty$ is referred to as \textit{the cusp}.

The low energy effective theory of the modulus $\tau$ is defined by a real K\"ahler potential $K(\tau,\bar\tau)$ and a
holomorphic superpotential $W(\tau)$.  The K\"ahler potential determines the geometry of field space; the metric is given by
\begin{align}
  K_{\tau\bar\tau} \;\equiv\; \partial_\tau \partial_{\bar\tau} K .
\end{align}
There is a redundancy in the definition of the K\"ahler potential and superpotential: two descriptions related by a K\"ahler transformation are physically equivalent
\begin{equation}
  K(\tau,\bar\tau) \;\to\; K(\tau,\bar\tau) + h(\tau)+\bar h(\bar\tau), 
  \qquad
  W(\tau) \;\to\; e^{-h(\tau)}\, W(\tau).
\end{equation}
The combination referred to as the \emph{generating functional}
\begin{equation}
  G(\tau,\bar\tau) \;\equiv\; K(\tau,\bar\tau) + \log |W(\tau)|^2 ,
\end{equation}
is invariant under these transformations and is the natural K\"ahler-invariant object.

We will assume supersymmetry is in the standard \emph{linearly} realized form where the scalar potential is 
\begin{equation}
  V_F \;=\; e^{G}\!\left(K^{\tau\bar\tau} G_\tau G_{\bar\tau} - 3\right)
  \;=\; e^{K}\!\left(K^{\tau\bar\tau} D_\tau W\, \overline{D_\tau W} - 3 |W|^2\right),
\end{equation}
where $K^{\tau\bar\tau}$ is the inverse of $K_{\tau\bar\tau}$, and the K\"ahler covariant derivative is given by $D_\tau W \;\equiv\; \partial_\tau W + (\partial_\tau K)\, W=W\cdot G_\tau$.
Supersymmetric vacua satisfy $D_\tau W=0$. SUSY breaking occurs when
\begin{equation}
  F^\tau \;=\; \,e^{G/2} K^{\tau\bar\tau} G_{\bar\tau}
      \;=\; \,e^{K/2} K^{\tau\bar\tau}\, \overline{D_\tau W}\;\neq\; 0,
\end{equation}
and the vacuum energy is set by the balance between $|F|^2$ and the universal $-3|W|^2$ term.

We will assume that we have a weak coupling infinte distance point when $\tau_2\gg 1$.  In such limits
the Kähler potential approaches
\be 
\label{kahlerleading} K\Big|_{\tau_2\gg 1}\sim -\kappa\,\text{log}\left(\frac{\tau-\Bar{\tau}}{2 i}\right).
\ee
 Indeed this is natural from the viewpoint of string theory as well as from the viewpoint of the distance conjecture, where $\tau_2\gg1$ is at infinite distance.
 
If we assume there are no other corrections and this is exact (which means we assume no corrections to the weak coupling limit) it implies that under the modular transformation, the Kähler potential transforms in the following way
\ba 
K\left(\tau,\Bar{\tau}\right)\rightarrow K\left(\tau,\Bar{\tau}\right)+ [\kappa\,\text{log}\left(c\,\tau+d\right)+c.c.].  
\ea
All observables of the theory are determined by the generating functional. Therefore, for the supergravity theory to be invariant under modular transformations, $G$ must be invariant.
 As a result, the superpotential must be a modular form of weight $-\kappa$
\be 
W(\tau)\rightarrow W(\tau) e^{-h(\tau)}=\frac{W(\tau)}{(c\,\tau+d)^\kappa}. 
\ee
More generally, the Kähler potential can be the logarithm of a \textit{real-analytic modular form} of weight $(k,k)$, which implies that the superpotential is a \textit{meromorphic modular form} of weight $-k$. \paragraph{Holomorphic modular forms.} 
A holomorphic modular form of weight $k$ is a holomorphic function such that 
\be
f(\gamma\tau)\;=\; (c\tau+d)^k\,f(\tau),
\ee
and $f$ is holomorphic at the cusp $\tau=i\infty$ (i.e. has a bounded $q$–expansion with $q=e^{2\pi i\tau}$). The graded ring of holomorphic modular forms for $\Gamma=SL(2,\mathbb{Z})$ is
\be
\mathcal{M}(\Gamma)=\mathbb{C}[E_4,E_6],
\ee
so every holomorphic modular form is a polynomial in the Eisenstein series $E_4$ and $E_6$. The holomorphic Eisenstein series of weight $k$ is defined in the following way
\be E_k(\tau)=\frac{1}{2}\sum_{(m,n)\neq (0,0)}\frac{1}{(n \tau+m)^k}. \ee
It has a q-expansion (Fourier expansion) of the form
\be 
E_k(\tau)=1-\frac{2k}{B_k} \sum_{n=1}^{\infty}\sigma_{k-1}(n) q^n, 
\ee
where $B_k$ is the $k$ th Bernoulli number, and $\sigma_k(n)=\sum_{d|n}d^k$ is the divisor function. \textit{Cusp forms} are those vanishing at the cusp, an example of which is $\Delta =\eta^{24}$ where Dedekind $\eta$ function is defined by 

\be \eta(\tau)=q^{1/24}\prod_{n\ge1}(1-q^n). \ee
$\eta$ function itself is not a regular modular form, but a weight $\frac{1}{2}$ modular form with multiplier system.  We discuss those later in the paper.
\paragraph{The elliptic $j$–invariant.}

The elliptic modular invariant is the weight–zero modular function
\be
j(\tau)=\frac{E_4(\tau)^3}{\Delta(\tau)}\,.
\ee
 In the $q \to 0$ limit, we have $j(\tau) \sim q^{-1}$ \cite{bruinier20081}.

\paragraph{Multiplier systems \& half-integral-weight forms} For a \textit{strict} modular form with trivial multiplier, under minus identity transformation of $SL(2,\mathbb{Z})$ modular forms are invariant, because it does not act on $\tau$.  However from the weight of a modular form we see that it picks up $(-1)^k$ which is consistent with the fact that $k\in 2\,\mathbb{Z}$.  In the superpotential it is allowed for $W$ to pick up extra phases under modular transformations.  This simply redefines which line bundle it is a section of.  Thus we must relax to a more general possibility for modular forms with a multiplier system.  The only allowed extra phases are 24-th roots of unity.
A canonical nontrivial multiplier comes from the Dedekind eta function
\be
\eta(\tau+1)=e^{\pi i/12}\,\eta(\tau),\quad
\eta(-1/\tau)=(-i\tau)^{1/2}\,\eta(\tau),
\ee
which is a weight \(1/2\) form with a multiplier system\footnote{Under $\gamma=\bigl(\begin{smallmatrix}a&b\\ c&d\end{smallmatrix}\bigr)\in SL_2(\mathbb Z)$, the Dedekind eta picks up a phase
$\eta(\gamma\tau)=\varepsilon(a,b,c,d)\,(c\tau+d)^{1/2}\eta(\tau)$ with
$\varepsilon(a,b,c,d)=e^{\pi i b/12}$ if $c=0,d=1$, and
$\varepsilon(a,b,c,d)=\exp\!\bigl[\pi i\bigl(\tfrac{a+d}{12c}-s(d,c)-\tfrac14\bigr)\bigr]$ for $c>0$,
where $s(h,k)=\sum_{n=1}^{k-1}\frac{n}{k}\bigl(\frac{hn}{k}-\lfloor \frac{hn}{k}\rfloor-\tfrac12\bigr)$.
}. By allowing a multiplier system, we can obtain half-integer-weight forms by multiplying an even-weight form by suitable powers of the Dedekind eta function, we have
\be
k \in \tfrac{1}{2}\mathbb{Z}.
\ee
Indeed modular forms with multiplier system are generated by the $(E_4,E_6,\eta)$, with the relation 
\be
\eta^{24}=\frac{1}{1728}(E_4^3-E_6^2).
\ee
So we only need powers of $\eta$ up to 24.

\paragraph{Real–analytic modular forms.}
A \emph{real–analytic modular form} of weight $(k,l)$ is a real–analytic function satisfying
\be
f(\gamma\tau)\;=\;(c\tau+d)^k\,(c\overline{\tau}+d)^{l}\,f(\tau).
\ee
 A prototype example of this is the real-analytic Eisenstein series of weight $(k,l)$, which is defined as \cite{2017arXiv170701230B,2017arXiv170803354B,2017arXiv171007912B}
\be 
\mathcal{E}_{k,l}(\tau,\bar{\tau})=\sum_{(m,n)\neq (0,0)}\frac{\tau-\Bar{\tau}}{2i (n\tau+m)^{k+1}(n\Bar{\tau}+m)^{l+1}}.
\ee
A special case of the above forms, relevant for constructing the Kähler potential, consists of those with equal holomorphic and anti-holomorphic weight, which we denote by
\be 
E_s(\tau,\bar{\tau},k)\equiv\sum_{(m,n)\neq (0,0)}\frac{(\tau-\Bar{\tau})^{s-k}}{(2i)^{s-k} \left|n\tau+m\right|^{2s}},
\ee
which is a real-analytic form of weight $(k,k)$. We use the following notation for the real-analytic Eisenstein function with weight zero
\ba 
&&E_s(\tau,\bar{\tau})\equiv E_s(\tau,\bar{\tau},0)=\sum_{(m,n)\neq (0,0)}\frac{\tau_2^s}{\left|n\tau+m\right|^{2s}}.
\ea
The real-analytic Eisenstein series of weight $(0,0)$ admits the q-expansion
\ba 
E_s(\tau,\bar{\tau}) &=& \Lambda(s) \tau_2^s + \Lambda(1-s) \tau_2^{1-s} \notag\\
&+&  \sum_{k=1}^{\infty} 4  k^{s-\sfrac{1}{2}} \sigma_{1-2s}(k) \sqrt{\tau_2} K_{s-\sfrac{1}{2}}(2\pi k \tau_2) \cos(2 \pi k \tau_1),\label{Eisenstein}
\ea  
where \( \Lambda(s) = \pi^{-s}\Gamma\left(s\right)\zeta(2s) \) is the completed zeta function \cite{Zagier1981EisensteinSA}. The real-analytic Eisenstein series
is an eigenfunction of the hyperbolic Laplacian, $\Delta E_s(\tau,\bar\tau) = s(1-s)\,E_s(\tau,\bar\tau)$.

\subsection{K\"ahler Potential as a Real-Analytic Form}
More generally, we will take the K\"ahler potential to be such that \(e^K\) transforms as a modular form of  weight \((k,k)\).  This implies that we can write the most general K\"ahler potential in the following way
\be e^{-K(\tau,\bar{\tau})}=\left(\frac{\tau-\bar{\tau}}{2 i}\right)^{k} F(j(\tau),\overline{j(\tau)}), \ee
where $j$ is the elliptic $j$-function.  $j$-function can be viewed as the $z$-coordinate on the moduli space viewed as a sphere. $F(j(\tau),\overline{j(\tau)})$ is a modular invariant, real positive function. Note that the weak coupling point $\tau_2\rightarrow \infty$, corresponds to $\tau_2\sim  \frac{1}{4\pi}\ {\rm log}(j(\tau)\overline{j(\tau)})\gg 1$

\paragraph{Asymptotic Behavior.} It is natural to assume that the Kähler potential has logarithmic behavior at the cusp as stated above; this assumption implies that the function $F$ exhibits the following asymptotic behavior 
\be F(j(\tau),\overline{j(\tau)})\simeq \left[\tfrac{1}{4\pi}\ {\rm log}(j(\tau)\overline{j(\tau)})\right]^{\alpha}+\text{sub-leading}\sim \tau_2^\alpha+\text{sub-leading}.\ee
Then the asymptotic K\"ahler potential is as follows
\be K_{asymp}=-\kappa \log\left(\frac{\tau-\bar{\tau}}{2 i}\right)\,;\; \kappa=k+\alpha, \ee
 In other words, {\it the modular weight does not by itself dictate the asymptotic behavior of } $K$; there are contributions from real-analytic terms in the Kähler potential. In more physical terms, the behavior at the cusp can differ from the naive weak coupling expectation. Recall that $k \in \tfrac{1}{2}\mathbb{Z}$ but $\alpha$ is in principle an arbitrary real number; to guarantee positivity of the metric asymptotically, we must have
\be
\kappa > 0 .
\ee
In the following we give examples of these real-analytic modifications when the Kähler potential includes the real-analytic Eisenstein series.  As we will see later, in many cases asymptotic behavior of the potential demanded by Swampland principles requires the stronger restriction: $\kappa \geq 1$.

\paragraph{Metric Positivity.} To ensure a well-defined theory, the metric derived from the Kähler potential must be positive definite. We have
\be
K=-k\,\log\!\left(\frac{\tau-\bar \tau}{2i}\right)-\log\!\left(F\big(j(\tau),\overline{j(\tau)}\big)\right),
\ee
therefore the metric reads
\be \label{metric}
K_{\tau\bar \tau}
= -\frac{k}{(\tau-\bar \tau)^2}
\;+\;
\frac{\partial_{\tau} F\big(j(\tau),\overline{j(\tau)}\big)\,\partial_{\bar{\tau}}F\big(j(\tau),\overline{j(\tau)}\big)}{F\big(j(\tau),\overline{j(\tau)}\big)^2}
\;-\;
\frac{\partial_{\tau} \partial_{\bar{\tau}}F\big(j(\tau),\overline{j(\tau)}\big)}{F\big(j(\tau),\overline{j(\tau)}\big)}, 
\ee
and the positivity condition is
\be
K_{\tau\bar{\tau}} \geq 0.
\ee

\paragraph{Metric positivity at the symmetry points.} Define the following local coordinate at a symmetry point $\tau_0$
\be
\delta \tau \equiv \tau - \tau_0,
\ee
Then the symmetry acts on this coordinate as
\be
\delta \tau \;\mapsto\; \zeta(\tau_0)\,\delta \tau\,; \qquad
\zeta(i) = -1, \quad \zeta\!\left(\omega\right) = e^{-\frac{2\pi i}{3}} ,
\ee
fixed by $S$, and $ST$ respectively. Therefore, at symmetry points the linear term in the expansion of $F(\tau,\bar{\tau})$ vanishes, and we have $\partial_{\tau} F=\partial_{\bar \tau} F=0$. Hence, at symmetry points the positive-definite term in \eqref{metric} vanishes and we obtain the positivity condition
\begin{equation}
K_{\tau\bar \tau}
= -\frac{k}{(\tau-\bar \tau)^2}
-\frac{\partial_{\tau} \partial_{\bar \tau}\, F\!\big(j(\tau),\overline{j(\tau)}\big)}{F\!\big(j(\tau),\overline{j(\tau)}\big)}
\Bigg|_{\tau=\tau_0}
\;\ge 0 .
\end{equation}
Equivalently,
\begin{equation}
\frac{\Delta F(j(\tau),\overline{j(\tau)})}{F(j(\tau),\overline{j(\tau)})}\Bigg|_{\tau=\tau_0}\;\le\; k,
\end{equation}
This gives an upper bound on the Laplacian of $F$ at symmetry points. 
While checking positivity directly can be difficult for arbitrary points, this bound at the symmetry points is simpler---and plausibly the most stringent near these points---because the positive-definite contribution vanishes at these points. 
Moreover, in cases where $F$ is a sum of eigenfunctions of the Laplacian---namely, the real-analytic Eisenstein series---the condition above is often much easier to analyze. 
We demonstrate this point in the examples that follow.

\subsubsection{Simple Examples of K\"ahler Potential}
A simple choice for the K\"ahler potential is logarithm of a single Eisenstein form of weight $(k,k)$
\ba 
&&K=\text{log}\,E_s(\tau,\bar{\tau},k).
\ea

\paragraph{Asymptotic Behavior.} At the cusp, the sum in the Eisenstein series is dominated by the zero mode of the Fourier expansion, and the Kähler potential approximates to  
\be 
K=-(k-s)\,\text{log}\left(\frac{\tau-\Bar{\tau}}{2i}\right),
\ee
We note that weight of the modular form does not necessarily dictate the asymptotic behavior of the K\"ahler potential, as anticipated.

\paragraph{Metric Positivity.} We can rewrite the Kähler potential in the following way
\be 
K = -k\,\log\left(\frac{\tau - \bar{\tau}}{2i}\right) + \log\,E_s(\tau, \bar{\tau}),
\ee
where $E_s(\tau, \bar{\tau})$ is the non-holomorphic Eisenstein series of weight zero, and is an eigen-function of the Laplacian
\be 
\Delta E_s(\tau,\bar\tau) = s(1-s)\,E_s(\tau,\bar\tau).
\ee  
We therefore find
\be \label{singlemetric}
K_{\tau\bar{\tau}} = \frac{s(1-s) - k}{(\tau - \bar{\tau})^2} 
- \frac{\partial_{\tau} E_s(\tau, \bar{\tau}) \partial_{\bar{\tau}} E_s(\tau, \bar{\tau})}{E_s(\tau, \bar{\tau})^2}.
\ee  
The first derivatives of the weight zero Eisenstein series are calculated as follows
\ba 
\partial_{\tau} E_s(\tau, \bar{\tau}) &=& s \sum_{(m_1, n_1) \neq (0,0)} 
\left( \frac{m_1 + n_1 \bar{\tau}}{m_1 + n_1 \tau} \right)
\left( \frac{(\tau - \bar{\tau})^{s-1}}{(2i)^s \left|n_1 \tau + m_1 \right|^{2s}} \right), \\
\partial_{\bar{\tau}} E_s(\tau, \bar{\tau}) &=& -s \sum_{(m_2, n_2) \neq (0,0)} 
\left( \frac{m_2 + n_2 \tau}{m_2 + n_2 \bar{\tau}} \right)
\left( \frac{(\tau - \bar{\tau})^{s-1}}{(2i)^s \left|n_2 \tau + m_2 \right|^{2s}} \right).
\ea
Therefore,
\be 
\partial_{\tau} E_s(\tau, \bar{\tau}) \partial_{\bar{\tau}} E_s(\tau, \bar{\tau})
= -\frac{s^2}{(\tau - \bar{\tau})^2} \left(\frac{\tau - \bar{\tau}}{2i}\right)^{2s}
\left| \sum_{(m, n) \neq (0,0)} 
\left( \frac{m + n \bar{\tau}}{m + n \tau} \right) 
\frac{1}{\left|n \tau + m \right|^{2s}} \right|^2.
\ee
We can now use the triangular inequality for the complex phasors $ 
\left|\sum_i z_i\right| \leq  \sum_i |z_i| $,
and get
\ba 
\partial_{\tau} E_s(\tau, \bar{\tau}) \partial_{\bar{\tau}} E_s(\tau, \bar{\tau}) 
&\leq&  -\frac{s^2}{(\tau - \bar{\tau})^2} 
\left[ \sum_{(m, n) \neq (0,0)} 
\frac{(\tau - \bar{\tau})^{s}}{(2i)^{s} \left|n \tau + m \right|^{2s}} \right]^2 \nn\\
&=& -\frac{s^2}{(\tau - \bar{\tau})^2} \left(E_s(\tau, \bar{\tau})\right)^2.\label{relation}
\ea
Therefore we have
\ba 
\frac{\partial_{\tau} E_s(\tau, \bar{\tau}) \partial_{\bar{\tau}} E_s(\tau, \bar{\tau}) }{E_s(\tau, \bar{\tau})^2}
\leq  -\frac{s^2}{(\tau - \bar{\tau})^2} .
\ea
This is the second term in Eq. \eqref{singlemetric}. Note that it exactly cancels the $s^2$ term in that equation. Therefore, the metric satisfies the inequality
\begin{equation}
K_{\tau\bar{\tau}} \ge \frac{s - k}{(\tau - \bar{\tau})^2} = \frac{k - s}{4 \tau_2^2}.
\end{equation}
The positivity requirement is satisfied if
\begin{equation}
k - s > 0.
\end{equation}

\subsubsection{IIB $\mathcal{N}=1$ Compactifications}
Setting axions and fluxes to zero, the coupling dependence of the large-volume perturbative Kähler potential of $\mathcal{N}=1$ IIB orientifold compactifications \cite{Becker:2002nn} is of the following form
\ba 
K &=& -\log \left(\frac{\tau - \bar{\tau}}{2i}\right) - 2 \log\left[V_E + 2 \zeta(3) \chi \frac{(\tau - \bar{\tau})^{\sfrac{3}{2}}}{(2i)^{\sfrac{3}{2}}} \right],
\ea
where $V_E$, and $\chi$ are the Einstein frame volume, and Euler character of the compact manifold respectively. It has been proposed in \cite{Green:1997tv,Grimm:2007xm} that the Kähler potential should be completed in those cases to the following object so as to exhibit the correct modular behavior
\ba
K &=& -\log \left(\frac{\tau - \bar{\tau}}{2i}\right) - 2 \log\left[V_E + \frac{\chi}{2} E_{\sfrac{3}{2}}(\tau,\bar{\tau})\right],\\
&=& -\log \left(\frac{\tau - \bar{\tau}}{2i}\right) - 2 \log\left[1 + \alpha \,E_{\sfrac{3}{2}}(\tau,\bar{\tau})\right]+c,
\ea
where $c \equiv -2 \log V_E$, and $\alpha \equiv \frac{\chi}{2 V_E} $\footnote{Note that the normalization of the real-analytic Eisenstein series here differs from \eqref{Eisenstein} by a factor of \(4\pi\).
}. 

\paragraph{Asymptotic Behavior.}
Near the cusp, the power–law part of the second term in the Eisenstein series, \eqref{Eisenstein}, dominates. Consequently, the K\"ahler potential has the asymptotic form
\be
K \simeq -4\log\!\left(\frac{\tau-\bar \tau}{2i}\right)
   = -4\log\!\big(\mathrm{Im}\,\tau\big).
\ee
This differs from the naive expectation based solely on the weight of the K\"ahler potential, reflecting that the Eisenstein series contributes a nontrivial modular–invariant piece whose power–law behavior overwhelms subleading terms near the cusp.

\paragraph{Metric Positivity.}  The K\"ahler metric is given by
\be 
K_{\tau\bar{\tau}} =-\frac{1}{(\tau - \bar{\tau})^2}
+\frac{2\alpha^2 \big(\partial_{\tau} E_{3/2}(\tau,\bar{\tau})\big)\big(\partial_{\bar{\tau}} E_{3/2}(\tau,\bar{\tau})\big)}{\big(1+\alpha\,E_{3/2}(\tau,\bar{\tau})\big)^2}
-\frac{2\alpha\,\partial_{\tau}\partial_{\bar{\tau}} E_{3/2}(\tau,\bar{\tau})}{1+\alpha\,E_{3/2}(\tau,\bar{\tau})}.
\label{kahler2}
\ee 
At the symmetry points, we have $\partial_{\tau} E_{\sfrac{3}{2}}(\tau,\bar{\tau})=\partial_{\bar \tau}E_{\sfrac{3}{2}}(\tau,\bar{\tau})=0$; therefore, the requirement of metric positivity at these points simplifies to
\ba K_{\tau\bar{\tau}}&=&-\frac{1}{(\tau-\bar \tau)^2}
-\frac{2\alpha\,\partial_{\tau}\partial_{\bar \tau}E_{3/2}(\tau,\bar \tau)}{1+\alpha\,E_{3/2}(\tau,\bar \tau)}\Bigg|_{\tau=\tau_0}\\
&=&\frac{\alpha\,E_{3/2}(\tau,\bar{\tau})-2}{2\,(\tau-\bar{\tau})^{2}\,\bigl(1+\alpha\,E_{3/2}(\tau,\bar{\tau})\bigr)}\Bigg|_{\tau=\tau_0}\geq 0,\ea
where, in the second line, we used the fact that the real-analytic Eisenstein series
is an eigenfunction of the hyperbolic Laplacian.
 
We carry out numerical calculations using the Fourier expansion of the Eisenstein series, given  in \eqref{Eisenstein}. At the symmetry points \(\tau=i\) and \(\tau=\omega\), the metric is positive definite for the following ranges of \(\alpha\)
\ba 
\tau=i\,:\;-0.111<\alpha<0.221, \\
\tau=\omega\,:\;-0.112<\alpha<0.225.
\ea
Furthermore, the metric blows up at points where \(e^{-K}=0\), which excludes all negative values of \(\alpha\).
These bounds translate to the range
\be
0 < \frac{\chi}{2V_E} < 0.221 .
\ee

\section{Superpotential and the Scalar Potential}
In Section~3.1, we discuss the most general half-integral-weight superpotential that can be constructed from modular forms and its pole structure. In Section~3.2, we constrain the superpotential and the asymptotic form of the K\"ahler potential according to the criteria provided by the Swampland program.
\subsection{Superpotential as a Meromorphic Form}
As noted above, the superpotential must be a meromorphic half-integral modular form of weight $-k$. For general $k$, every meromorphic modular form of weight $-k$ can be written as
\ba \label{superpot}
W(\tau)=R(j(\tau))\,E_4(\tau)^{\alpha}\,E_6(\tau)^{\beta}\, \eta(\tau)^{m},
\ea
where we impose \(4\alpha + 6\beta + \frac{m}{2} = -k\) to ensure the correct modular weight.  We can choose $0\leq m< 24$. The function \(R\) is a rational function of the \(j\)-function.  To see this note that if we divide $W$ by any meromorphic object of the right modular weight and transformation property (and in particular by $E_4(\tau)^{\alpha}\,E_6(\tau)^{\beta}\, \eta(\tau)^{m}$) we get a meromorphic modular invariant object and that is generated by $j$ and so it is a rational function $R(j)$.

Depending on how the modular forms are combined, the superpotential can develop \emph{poles} and \emph{zeros}. Note that \(E_4(\tau)\) has a simple zero at \(\tau=\omega\), \(E_6(\tau)\) has a simple zero at \(\tau=i\), and the Dedekind \(\eta(\tau)\) has no zeros in the upper half-plane, vanishing only at the cusp \footnote{By modular invariance, the zeros/poles of a weight-\(k\) modular form \(W\) satisfy the valence formula \(\sum_{P}\frac{1}{n_P}\operatorname{ord}_P(W)+\operatorname{ord}_\infty(W)=k/12\), where \(n_P=2,3\) at the elliptic points \(i,\ \omega=e^{2\pi i/3}\) and \(n_P=1\) otherwise; hence any added pole must be compensated by added zeros.
}.
\paragraph{Regularity at the cusp.}  Physically, poles in the interior of the fundamental domain correspond to loci in the theory where some light fields become massless; however, poles at the cusp are disallowed because they lead to a blow-up in the asymptotic behavior of the superpotential, and hence of the scalar potential. At weak coupling, we assume $V \to 0$ forbidding this behavior. Using the $q$-expansions of the modular forms involved, we find that as $q \to 0$ the superpotential behaves as follows
\be
W(\tau)\sim R(q^{-1})\,q^{\frac{m}{24}}.
\ee
If \( R(x) = \frac{P(x)}{Q(x)} \), where \( P(x) \) and \( Q(x) \) are polynomials of degrees \( a \) and \( b \), respectively, then the asymptotic behavior of the superpotential is as follows
\be
W(\tau)\sim q^{\frac{m}{24}-(a-b)}.
\ee
Avoiding an asymptotic blow-up of $W$ (and hence of $V$) requires
\be \label{condition1}
\frac{m}{24}-(a-b)\geq 0.
\ee

\paragraph{Regularity at the symmetry points.} We can also determine the regularity conditions at the symmetry points. $E_4$ has a simple zero at the point $\tau_0=\omega$: $ E_4(\tau)\sim \tau-\omega$. $E_6$ has a simple zero at the point $\tau_0=i$: $E_6(\tau)\sim \tau-i$.  Let us denote the order of the rational function $R$ at $j=j(a)$ by $v_a(R)$, meaning that at $j=j(a)$ the rational function behaves as
\be R(j)\sim (j-j(a))^{v_a(R)}. \ee
We have $j(\omega)=0$ and $j(i)=1728$, and
\be j(\tau)-1728\Big|_{\tau=i}\sim (\tau-i)^2\;,\qquad j(\tau)\Big|_{\tau=\omega}\sim (\tau-\omega)^3. \ee 
Therefore, the symmetry point $\tau_0=i$ is regular if \be \label{nopolei}\beta+2\,v_{i}(R)\geq 0, \ee and the symmetry point $\tau_0=\omega$ is regular if \be \label{nopoleomega}\alpha+3\,v_{\omega}(R)\geq 0. \ee
We will return to these conditions when analyzing the structure of the vacua at the symmetry points.

\subsection{Asymptotic Behavior of the Scalar Potential}
We now discuss asymptotic behavior of the scalar potentials that arise in four-dimensional ${\cal N}=1$ supergravity theories with modular symmetry, using the superpotential constructed in the previous section together with the following \textit{asymptotic} K\"ahler potential
\begin{equation}
K \;=\; -\kappa\,\log\!\left(\frac{\tau-\bar{\tau}}{2 i}\right).
\end{equation}
Not all scalar potentials can descend from a theory of quantum gravity. The Swampland program imposes two important conditions on the asymptotic behavior of the scalar potential: \emph{(i) Vanishing asymptotics, (ii) Slope bound: $\lvert\nabla_\phi V\rvert/V \geq \sqrt{2}$.}

We will use these criteria to constrain the superpotential and the asymptotic form of the K\"ahler potential. In our setup, the leading behavior of the superpotential in the asymptotic ($q\rightarrow 0$) limit is given by
\begin{equation}
W(\tau)\sim q^{c_1}(1+c_2 q+...).
\end{equation}

\paragraph{(i) Vanishing asymptotics.}
In the asymptotic weak coupling regime, the potential should tend to zero, and any blow up of $V$ is forbidden. Avoiding an asymptotic blow-up of $W$ (and hence of $V$) requires
\be \label{condition1}
\boxed{c_1\geq 0}.
\ee

For $c_1>0$, the potential exhibits double-exponential decay, which in principle is allowed and signifies that all perturbative corrections to $V$ vanish. In terms of the canonically normalized field, $\phi=\sqrt{\frac{\kappa}{2}}\log\tau_2$, we find
\be
\frac{\nabla_{\phi} V}{V} \simeq -4\pi c_1\,\sqrt{\frac{2}{\kappa}}\, e^{\sqrt{\frac{2}{\kappa}}\phi}.
\ee
Double-exponential decay is reasonable and may be implied by instanton effects in gauge theories. For $c_1=0$, the potential exhibits exponential decay which we now turn to. 
\paragraph{(ii) Slope bound.}
Swampland ideas constrain the asymptotic slope of the scalar potential in four dimensions as follows \cite{Bedroya:2019snp, Rudelius:2021oaz}
\begin{equation}\label{nabla}
\frac{|\nabla V|}{V}\;\ge\;\frac{2}{\sqrt{d-2}}=\sqrt{2}\,,
\end{equation}
favoring exponential decay.  With $c_1=0$, the leading $q\rightarrow 0$ behavior of the superpotential is
\be
W\simeq 1+c_2\,q+\dots,
\ee
and the scalar potential can exhibit two kinds of asymptotic behavior. For $\kappa\neq 3$,
\be
\frac{\nabla_{\phi} V}{V} = -\sqrt{2\,\kappa},
\ee
so the potential decays exponentially and satisfies the swampland slope bound for
\be \label{condition2}
\boxed{\kappa\geq 1}.
\ee
For $\kappa=3$, the leading piece cancels  and one obtains double-exponential decay
\be
\frac{\nabla_{\phi} V}{V}\simeq -2\pi\sqrt{\frac{2}{3}}\, e^{\sqrt{\frac{2}{3}}\phi}.
\ee
This is similar to the co-called \textit{no-scale} behavior.


\section{Structure of Vacua at the Symmetry Points from Modularity}
\subsection{Symmetries of the Scalar Potential}
We investigate the behavior of the scalar potential at the symmetry points $\tau_0=i,\omega$, which are stabilized by $S$ and $ST$, respectively.  We will assume that at these points $V$ is regular.  We return below to examining when regularity is a good assumption.

Recall that the scalar potential is given by
\be V=e^K\left(K^{\tau \bar{\tau}} D_{\tau}W D_{\bar{\tau}}\bar{W}-3|W|^2\right), \ee
 $W$ is a holomorphic form of weight $-k$. While $\partial_{\tau} W$ is not a form, we show that due to the transformation property of the K\"ahler potential, $D_{\tau} W$ transforms similar to a modular form of weight $2-k$.

\paragraph{Transformation Property of $D_{\tau}W$ under SL(2,Z).}
The covariant derivative of the superpotential is
\be D_{\tau}\,W=\partial_{\tau}\,W+(\partial_{\tau}\,K) W, \ee
where $W$ is a holomorphic form of weight $-k$, and $e^K$ is a real-analytic form of weight $(k,k)$. $D_{\tau} W$ is related to the invariant quantity, $G=K+\text{log}\,\bar{W}W$, through the following relation
\be D_{\tau} W=\frac{\partial_{\tau}\left(e^G\right)}{e^K \bar{W}}.\ee
Under the modular transformation, we have
\be D_{\tau} W\rightarrow (D_{\tau} W)'=(c \tau+d)^{2-k}\frac{\partial_{\tau}\left(e^G\right)}{e^K\bar{W}}=(c \tau+d)^{2-k} \,D_{\tau} W, \ee
where we have used the fact that $\sfrac{\partial \tau}{\partial \tau'}=(c \tau+d)^2$ and that $G$ is invariant under modular transformations. Therefore, $D_{\tau} W$ transforms as a form of weight $2-k$.

\paragraph{Criticality at the Symmetry Points.}
 Under modular transformations, the derivative of the scalar potential transforms as
\be \partial_{\tau} V\rightarrow (c \tau+d)^2 \partial_{\tau} V. \ee
At the symmetric points $\tau_0=i,\,\omega$, the transformations $S,\,ST$ that stabilize them, respectively, give
\ba &&\tau=i;\;\;S:\, \partial_{\tau} V\rightarrow -\partial_{\tau} V, \\
&&\tau=\omega;\,ST:\, \partial_{\tau} V\rightarrow e^{\frac{2 \pi i}{3}}\partial_{\tau} V.\ea
Therefore we have
\be \partial_{\tau} V\Big|_{\tau_0}=0, \ee
and criticality at these symmetric points is guaranteed by modular properties. In what follows, we determine the sign of the scalar potential at these critical points—i.e., whether the vacuum is de Sitter, supersymmetric anti–de Sitter, or supersymmetric Minkowski, which we get for $(D_\tau W\neq 0, W=0)$, $(D_\tau W= 0, W\neq0)$, and $(D_\tau W=W=0)$ respectively. 

\paragraph{Regularity assumption.}
Above we have been assuming that $V,W$ are regular at the symmetry points. Here we examine necessary and sufficient conditions for this to be true.
Regularity at the symmetry points requires (see \eqref{nopolei} for $\tau=i$ and \eqref{nopoleomega} for $\tau=\omega$)
\begin{align}
&\tau_0=i:\quad \beta + 2\,v_{i}\ge 0, \label{eq1}\\
&\tau_0=\omega:\quad \alpha + 3\,v_{\omega}\ge 0. \label{eq2}
\end{align}
To obtain the correct weight we impose
\[
4\alpha+6\beta=-k-\frac{m}{2}.
\]
In particular, if \eqref{eq1} and \eqref{eq2} hold, then both points are regular; if either \eqref{eq1} or \eqref{eq2} fails, the corresponding point is a pole.

\subsection{No Multiplier System; Forms of Even Weight}
In this section we will assume symmetry points are regular and consider the case in which there is no multiplier system and the forms are of even weight. This is equivalent to setting $m=0 \;\text{mod}\; 24$ in the expansion \eqref{superpot}. The symmetry at the symmetry points constrains the modular forms $W$ and $D_{\tau}W$ in the following way. At the point $\tau=i$, after transformation under $S$, we have 
\ba &&W(i)\rightarrow (i)^{-k}\,W(i),\\
&& D_{\tau} W(i)\rightarrow (i)^{2-k}\,D_{\tau} W(i).\ea
Thus, assuming regularity at $\tau=i$, we obtain
\ba &&W(i)=0\,\text{ for }\; k\neq 4\,n ,\\
&& D_{\tau} W(i)=0\,\text{ for }\; k-2\neq 4\,n, \ea
where $n\in \mathbb{Z}$. Similarly, at the point $\tau=\omega$, after transformation under $ST$, we have
\ba &&W(\omega)\rightarrow \left(e^{\frac{i \pi}{3}}\right)^{-k}\,W(\omega),\\
&& D_{\tau} W(\omega)\rightarrow \left(e^{\frac{i \pi}{3}}\right)^{2-k}\,D_{\tau} W(\omega).\ea
Thus, assuming regularity at $\tau=\omega$, we have
\ba &&W(\omega)=0\,\text{ for }\; k\neq 6\,n ,\\
&& D_{\tau} W(\omega)=0\,\text{ for }\; k-2\neq 6\,n. \ea

Note that the nature of vacua at symmetry points depends only on $k$ mod 12.  We list the vacua for different even values of $k$ mod 12 in tables \ref{tab:even-i}, and \ref{tab:even-omega}.   We recall that weight of the modular form is not dictated by the asymptotic behavior of the K\"ahler potential.
\begin{table}[h]
\centering
\begin{tabular}{|c|c|c|c|c|c|c|c|c|c|c|}
\hline
$k$ & 2 & 4 & 6 & 8 & 10 & 12  \\ \hline
$W$ & 0 & $\neq 0$  & 0  & $\neq 0$  & 0 & $\neq 0$ \\ \hline
$D_{\tau}W$ & $\neq 0$  & 0  & $\neq 0$  & 0  & $\neq 0$ & 0 \\ \hline
$V$  & dS  & AdS  & dS  & AdS  & dS & AdS \\ \hline
\end{tabular}
\caption{Vacua at the symmetry point \(\tau=i\) for even modular weight \(k\).}
\label{tab:even-i}

\end{table}
\begin{table}[h]
\centering
\begin{tabular}{|c|c|c|c|c|c|c|c|c|c|c|}
\hline
$k$ & 2 & 4 & 6 & 8 & 10 & 12 \\ \hline
$W$ & 0  & 0  & $\neq 0$  & 0  & 0 & $\neq 0$ \\ \hline
$D_{\tau}W$ & $\neq 0$ & 0  & 0  & $\neq 0$  & 0 & 0 \\ \hline
$V$  & dS  & Mink & AdS  & dS & Mink & AdS \\ \hline
\end{tabular}
\caption{Vacua at the symmetry point \(\tau=\omega\) for even modular weight \(k\).}
\label{tab:even-omega}
\end{table}

It is remarkable that simply the knowledge of modular weight, without any additional input from dynamics of the theory, dictates the generic nature of vacua at symmetry points!  The only assumption in this derivation is the regularity of the potential at these points. In other words, we can predict that at these points there are either massless fields (signaling break down of EFT and the lack of regularity) or predict the nature of vacua.

\subsection{Incorporating a Multiplier System}
As mentioned before, we can obtain forms of odd or half-integer weight by including $\eta^{m}$ factors, since $W$ can have a multiplier system. We can do a Kahler tranformation an equivalent theory with a form of weight $-k'=-k-\sfrac{m}{2}$, $k'=2n,\, n\in \mathbb{Z}$,  for the superpotential in the following way
\be \label{hintneg} W'_{-k'}(\tau)=W_{-k}(\tau)  \eta(\tau)^{-m}\quad K'=K+m\ {\rm log}(\eta \, \bar \eta). \ee
Note that this does not change the regularity properties of $W$ in the interior, but makes it have even modular weight as was the case without the $\eta$ factors.  We thus obtain the same structure of vacua as in tables~\ref{tab:even-i} and~\ref{tab:even-omega}, with $k$ replaced by $k'$ . In summary, a plethora of dS, Minkowski, or AdS vacua occurs at the symmetry points. These can be minima, maxima, or saddle points of the scalar potential.

\section{Discussion}
In this paper, we have shown that in four-dimensional $\mathcal{N}=1$ supergravity with modular symmetry, the symmetry points $\tau=i$ and $\tau=\omega$ are always critical points of the scalar potential (assuming no additional massless fields at these points), and that the \emph{sign} of the vacuum energy there --- $\mathrm{dS}$, $\mathrm{AdS}$, or Minkowski --- is fixed by the modular weight of the superpotential (after getting rid of the multiplier part). Our classification thus provides a symmetry based principle to pinpoint existence and nature of extrema without detailed knowledge of $W$ or the full non-perturbative $K$.

We also studied the asymptotic behavior of the scalar potential. At large $\tau_2$, the behavior is controlled by an \emph{asymptotic} Kähler slope $\kappa$, which can deviate from its naive tree-level value due to real-analytic modular contributions and yields exponential or double exponential decay. We exclude double-exponentially growing potentials near the cusp by imposing constraints on the superpotential. Also using swampland bounds on the slope of the scalar potential, we place bounds on the asymptotic Kähler potential.

It would be interesting to draw conclusions about the sign of the Hessian at the elliptic points directly from symmetry considerations (perhaps combined with other features) and thereby make statements about (in)stability.  One could also extend the analysis to congruence subgroups and multi-modulus moduli spaces, where additional elliptic points further enrich the vacuum structure. When there are gauge fields, it may be worthwhile analyzing the gauge kinetic function, and hence the gauge coupling, from the modular perspective.
Finally, it would be interesting to actually find examples of theories with ${\cal N}=1$ modular geometries realized in a consistent string landscape, as none is currently known!
\\
\section*{Acknowledgments}

We would like to thank Damian van de Heisteeg for valuable discussions. AM thanks the CMSA for the Workshop on Symmetries and Gravity and providing a stimulating environment during the workshop, while this work was in progress.

This work is supported in part by a grant from the Simons Foundation (602883, CV) and the DellaPietra Foundation.

\bibliographystyle{jhep}
\bibliography{references}
\end{document}